\documentclass{article}
\makeatletter
\@addtoreset{equation}{section}

\makeatother

\begin{document}

\begin{flushright}
Oct 2006

KUNS-2042
\end{flushright}

\begin{center}

\vspace{5cm}

{\LARGE 
\begin{center}
Tachyon Condensation with $B$-field
\end{center}
}

\vspace{2cm}

Takao Suyama \footnote{e-mail address: suyama@gauge.scphys.kyoto-u.ac.jp}

\vspace{1cm}

{\it Department of Physics, Kyoto University,}

{\it Kitashirakawa, Kyoto 606-8502, Japan}

\vspace{4cm}

{\bf Abstract} 

\end{center}

We discuss classical solutions of a graviton-dilaton-$B_{\mu\nu}$-tachyon system. 
Both constant tachyon solutions, including $AdS_3$ solutions, 
and space-dependent tachyon solutions are investigated, and their 
possible implications to closed string tachyon condensation are argued. 
The stability issue of the $AdS_3$ solutions is also discussed.

\newpage

\vspace{1cm}

\section{Introduction}

\vspace{5mm}

Closed string tachyon condensation has been discussed recently \cite{review1}\cite{review2}. 
However, the understanding of this phenomenon is still under progress, compared with the open string 
counterpart \cite{open}. 
A reason for the difficulty would be the absence of a useful tool to investigate closed string tachyon 
condensation. 
It is desired that there would be a non-perturbative framework for this purpose. 
Closed string field theory was applied to the tachyon condensation in string theory on 
orbifolds, and preliminary results were obtained \cite{SFT}, 
but the analysis would be more complicated than 
that in the open string case. 
To make a modest step forward, a classical gravitational theory coupled to a tachyon was discussed 
as a leading order approximation to the closed string field theory 
\cite{YZ}\cite{YZ2}\cite{fixedpoint}\cite{Bergman}. 
In \cite{YZ}\cite{fixedpoint}\cite{Bergman}, 
a theory of graviton, dilaton and tachyon was discussed, and its classical time evolution 
was investigated. 

In this paper, we investigate a similar theory, with NS-NS $B$-field also taken into account. 
By the presence of the $B$-field, an $AdS_3$ background is allowed as a classical solution, in addition 
to the well-known linear dilaton solution. 
Note that similar $AdS_3$ solutions in the context of gauged supergravity were already found in 
\cite{gaugedSUGRA}\cite{gaugedSUGRA2}, and the BTZ black hole in almost the same action was discussed 
in \cite{HW}. 
We also investigate solutions in which the tachyon field varies along a spatial direction. 
These solutions would provide information on the backreaction of turning on the vev of the tachyon 
to the other background fields. 
The stability issue of the $AdS_3$ background is also discussed, focusing on the presence of 
localized tachyons. 

This paper is organized as follows. 
In section \ref{solution}, 
we investigate classical solutions of a graviton-dilaton-$B_{\mu\nu}$-tachyon system. 
Both constant tachyon solutions and space-dependent tachyon solutions are considered. 
In section \ref{stability}, 
the presence of tachyons in bosonic string theory on $AdS_3$ backgrounds is analyzed. 
Section \ref{discuss} is devoted to discussion.

\vspace{1cm}

\section{Classical solutions  \label{solution}}

We consider classical solutions of the following action, 
\begin{equation}
S = \frac1{2\kappa^2}\int d^Dx\sqrt{-g}\ e^{-2\Phi}\Bigl[ R+4(\nabla\Phi)^2-\frac1{12}H^2-(\nabla T)^2
 -V(T) \Bigr]. 
    \label{action}
\end{equation}
The same action with $B_{\mu\nu}=0$ was discussed in \cite{YZ}\cite{fixedpoint}\cite{Bergman}. 
If we regard this action as a low energy effective theory of a string theory, it is understood that 
there are other spatial directions, for example $26-D$ directions 
for bosonic string, which are compactified on 
a manifold $M$. 
The tachyon $T$ may come from a relevant operator of a CFT describing $M$. 
We assume that $V(T)$ has a maximum at $T=T_{max}$ and a minimum at $T=T_{min}>T_{max}$, and also 
assume $V(T)<0$ for $T_{max}\le T\le T_{min}$. 

The equations of motion are as follows, 
\begin{eqnarray}
R_{\mu\nu}+2\nabla_\mu\nabla_\nu\Phi-\frac14H_{\lambda\rho\mu}H^{\lambda\rho}{}_\nu
 -\nabla_\mu T\nabla_\nu T &=& 0, \\
\nabla^2\Phi-2(\nabla\Phi)^2-V(T)+\frac1{12}H^2 &=& 0, \\
\partial_\rho(\sqrt{-g}\ e^{-2\Phi}H^{\rho\mu\nu}) &=& 0, 
   \label{eomB} \\
\nabla^2 T-2\nabla\Phi\cdot\nabla T-V'(T) &=& 0, 
\end{eqnarray}
where $V'(T)=\frac{dV(T)}{dT}$. 

\vspace{5mm}

\subsection{Vacuum solutions}

\vspace{5mm}

First, let us consider vacuum solutions. 
The ``vacuum'' means that the tachyon field $T$ is constant. 
The equation of motion of $T$ implies that $V'(T)=0$, so we choose $T=T_*$ where $T_*=T_{max}$ or 
$T_*=T_{min}$. 

When $H_{\mu\nu\rho}=0$, it is well-known that a linear dilaton background
\begin{equation}
g_{\mu\nu}=\eta_{\mu\nu}, \hspace{5mm} \Phi = \Phi_0\pm\sqrt{-\frac12V(T_*)}\ x, 
\end{equation}
where $x$ is a spatial coordinate, is a solution of the equations of motion. 

There is also a solution when $H_{\mu\nu\rho}\ne0$. 
Let us make the following ansatz, 
\begin{eqnarray}
ds^2 &=& \hat{g}_{\alpha\beta}(x)dx^\alpha dx^\beta+\bar{g}_{kl}(y)dy^kdy^l, 
   \label{ansatz1} \\
H^{012} &=& \frac1{\sqrt{-\hat{g}}}h(x), 
   \label{ansatz2}
\end{eqnarray}
where $\alpha,\beta=0,1,2$ and $k,l=3,\cdots,D-1$, and the other components of $H^{\mu\nu\rho}$ 
are set to be zero. 
Then, there is a solution with a {\it constant dilaton}, 
\begin{equation}
\hat{R}_{\alpha\beta} = -\frac12e^{4\Phi_0}\hat{g}_{\alpha\beta}, \hspace{5mm} 
\bar{R}_{kl} = 0, \hspace{5mm} 
h = \pm e^{2\Phi_0} = \pm\sqrt{-2V(T_*)}. 
\end{equation}
A typical solution is $AdS_3\times {\bf R}^{D-3}$. 
The radius of $AdS_3$ is determined by the depth of the potential $V(T)$. 
The radius is small as the potential is deep. 
An explicit form of the $AdS_3$ solution is 
\begin{equation}
ds^2 = R^2\Bigl[ r^2(-dt^2+dy^2)+\frac{dr^2}{r^2} \Bigr], 
   \label{AdSmetric}
\end{equation}
where $R^2=\frac2{-V(T_*)}$. 
Since the dilaton is constant in this solution, whose value can be chosen arbitrarily, 
the perturbative analysis of string theory on this solution 
is reliable.

\vspace{5mm}

\subsection{Interpolating solutions}

\vspace{5mm}

Next, we investigate classical solutions which interpolate the above vacuum solutions, that is, 
solutions which vary along a spatial direction, say $x$, and which approach one of the above vacuum 
solutions in the limit $x\to\pm\infty$. 
The relevance of such solutions is as follows. 
Without loss of generality, one can set $T_{max}=0$. 
This maximum can be regarded as a perturbative 
vacuum of a string theory which has a tachyon in the mass spectrum, 
after choosing a background metric etc. 
Then by turning on the vev of $T$ with its backreaction taking into account, 
one can in principle find 
which background would be realized at $T=T_{min}\ne0$. 
This procedure would be equivalent to finding a classical solution, varying along the $x$-direction, 
in which $T(x)$ interpolates $T=0$ and $T=T_{min}$. 

This idea is in analogy with the open string tachyon condensation. 
Tachyon field on an unstable D-brane can take a kink solution which asymptotically approaches a 
minimum of a tachyon potential. 
In the asymptotic region the D-brane disappears, while around the center of the kink a lower dimensional 
D-brane remains. 
This suggests that a homogeneous tachyon condensation would result in a complete decay of the D-brane. 

It should be noted that the above analysis might not imply that some background would be realized 
after a {\it dynamical} process of tachyon condensation, which should be discussed in another way. 

\vspace{5mm}

To find the desired solutions, we again employ the ansatz (\ref{ansatz1})(\ref{ansatz2}) with 
the following more detailed form of the metric, 
\begin{equation}
\hat{g}_{\alpha\beta}dx^\alpha dx^\beta = -e^{-2A(x)}dt^2+dx^2+e^{-2B(x)}dy^2, 
\end{equation}
and we also assume that $\Phi$ and $T$ depend only on $x$. 

The equation of motion (\ref{eomB}) implies, after a suitable shift of $\Phi$, 
\begin{equation}
h = \pm \epsilon e^{2\Phi(x)}, 
\end{equation}
where $\epsilon$ is 0 or 1, depending on whether the integration constant is zero or not. 
Solutions with $\epsilon=0$ have the vanishing $H$-flux. 

Now the equations of motion reduce to 
\begin{eqnarray}
A''+B''-(A')^2-(B')^2+2\Phi''+\frac12\epsilon e^{4\Phi}-(T')^2 &=& 0, \\
A''-A'(A'+B')-2A'\Phi'+\frac12\epsilon e^{4\Phi} &=& 0, \\
B''-B'(A'+B')-2B'\Phi'+\frac12\epsilon e^{4\Phi} &=& 0, \\
\Phi''-(A'+B')\Phi'-2(\Phi')^2-V(T)-\frac12\epsilon e^{4\Phi} &=& 0, \label{eomPhi} \\
T''-(A'+B')T'-2\Phi'T'-V'(T) &=& 0, 
\end{eqnarray}
where the prime indicates the derivative with respect to $x$. 

This system of equations looks complicated, but this becomes tractable when 
\begin{equation}
K=A'+B'+2\Phi'
\end{equation}
is introduced. 
It is easy to show that $K'=K^2+2V(T)$ by using the equations of motion. 
Therefore, we can solve the following two equations 
\begin{eqnarray}
T''-KT'-V'(T) &=& 0, 
   \label{T} \\
K'-K^2-2V(T) &=& 0, 
   \label{K}
\end{eqnarray}
first, and then solve the other equations with $T$ and $K$ regarded as a given function of $x$. 

\vspace{5mm}

The qualitative behavior of $T(x)$ and $K(x)$ can be easily deduced from the equations 
(\ref{T})(\ref{K}). 
The equation (\ref{T}) is the familiar equation of motion of a point particle in the potential 
$-V(T)$ with a position-dependent ``friction'' term. 
A solution we would like to find should behave asymptotically, 
\begin{equation}
T(x) \to \left\{
\begin{array}{cc}
T_{min}, & (x\to-\infty) \\ T_{max}. & (x\to+\infty)
\end{array}
\right.
\end{equation}
To obtain the above behavior for $x\to+\infty$, $K$ must be negative in this limit. 
Indeed, $K=-\sqrt{-2V(T_{max})}$ is the stable fixed point of $K$ when $T\sim T_{max}$, 
and therefore such a solution can exist. 
The solution may oscillate around $T=T_{max}$, depending on the value of $K(+\infty)$. 
It can be shown that the oscillation occurs when $4V''(T_{max})<-K(+\infty)^2$. 
This condition exactly coincides with the condition for the field $T$ to be really tachyonic around 
$T=T_{max}$, that is, 
its mass$^2$ is below the Breitenlohner-Freedman bound \cite{BF1}\cite{BF2} for the 
$AdS_3$ space (\ref{AdSmetric}). 

We would like to find $K(x)$ which behaves as follows, 
\begin{equation}
K(x) \to \left\{
\begin{array}{cc}
-\sqrt{-2V(T_{min})}, & (x\to-\infty) \\ -\sqrt{-2V(T_{max})}. & (x\to+\infty)
\end{array}
\right. 
   \label{asymptoticK}
\end{equation}
In this solution, $K$ 
starts at a stable fixed point, and end at another stable fixed point which exists 
since $T$ varies along $x$, so such a solution may exist. 
Note that there might exist a solution in which $K(-\infty)=+\sqrt{-2V(T_{min})}$. 
As we will discuss in section \ref{discuss}, this solution has a peculiar feature in view of RG flow, 
and therefore, it will not be 
investigated here. 
Another reason for ignoring such solutions is that, if $K>0$ at some range of $x$, then $T$ may 
be accelerated too much, and its motion may be affected by the detailed shape of $V(T)$ in a wide range 
of $T$. 
On the other hand, the qualitative behavior of solutions with (\ref{asymptoticK}) depends almost only 
on the existence of a minimum and a maximum of $V(T)$. 

In summary, the equations (\ref{T})(\ref{K}) admit a solution with the desired asymptotic behavior. 
The existence of such a solution can be easily confirmed by the numerical integration of the equations. 

\vspace{5mm}

Next, let us investigate the behavior of $\Phi$ with $T$ and $K$ given above. 
When $\epsilon=0$, (\ref{eomPhi}) can be easily integrated once, and the result is, 
\begin{equation}
\Phi'(x) = e^{{\cal K}(x)}\int_{-\infty}^x d\xi\ e^{-{\cal K}(\xi)}V(T(\xi))+Ce^{{\cal K}(x)}, 
   \label{solnPhi}
\end{equation}
where ${\cal K}'=K$. 
This solution behaves $\Phi'\to Ce^{K(-\infty)x}$ in $x\to-\infty$, but this corresponds neither 
a linear dilaton solution nor $AdS_3$ solution. 
Therefore we set $C=0$. 
Then $\Phi'$ behaves as 
\begin{equation}
\Phi'(x) \to -\frac{V(T(\pm\infty))}{K(\pm\infty)} = -\sqrt{-\frac12V(T(\pm\infty))} \hspace{5mm}
(x\to\pm\infty). 
\end{equation}
This shows that 
(\ref{solnPhi}) corresponds to a solution which interpolates two linear dilaton solutions. 
Note that the remaining equations imply $A'=B'=0$. 

We then consider $\epsilon=1$ case. 
In this case, the equation for $\Phi$ is complicated, so we only discuss its asymptotic behavior 
which is obtained by solving 
\begin{equation}
\Phi''-K_\pm\Phi'-V_\pm-\frac12e^{4\Phi} = 0, 
\end{equation}
where $K_\pm=K(\pm\infty)$ and $V_\pm=V(T(\pm\infty))=-\frac12K_\pm^2$. 
This equation can be rewritten as follows, 
\begin{equation}
\frac{d^2X_\pm}{d\rho^2} = e^{2X_\pm}, 
   \label{X}
\end{equation}
where 
\begin{equation}
X_\pm = 2\Phi-K_\pm x-\log|K_\pm|, \hspace{5mm} \rho = e^{K_\pm x}. 
\end{equation}
The solution is 
\begin{equation}
e^{-X_\pm(\rho)} = \frac1{\sqrt{2E_\pm}}\sinh(\sqrt{2E_\pm}(\rho+\rho_\pm)), 
\end{equation}
where $E_\pm\ge0$ and $\rho_\pm\ge0$ are integration constants. 

In $x\to+\infty$, corresponding to $\rho\to0$, $\Phi$ behaves as follows, 
\begin{equation}
e^{2\Phi} \to \left\{ 
\begin{array}{lc}
|K_+| & (\rho_+=0), \\ 
|K_+|e^{X_+(0)}e^{K_+x} & (\rho_+\ne0). 
\end{array}
\right.
\end{equation}
The $\rho_+=0$ solution approaches the $AdS_3$ solution, while the $\rho_+\ne0$ solution 
approaches the linear dilaton solution. 

On the other hand, in $x\to-\infty$, $\Phi$ behaves as follows, 
\begin{equation}
e^{2\Phi} \to \left\{ 
\begin{array}{lc}
|K_-|\sqrt{2E_-}\rho\ e^{-\sqrt{2E_-}(\rho+\rho_-)} & (E_->0), \\ 
|K_-| & (E_-=0). 
\end{array}
\right.
\end{equation}
The $E_-=0$ solution approaches the $AdS_3$ solution, but the $E_->0$ solution does not approach 
neither a linear dilaton nor $AdS_3$. 

\vspace{5mm}

We have found that there is an asymmetry in the existence of the interpolating solutions. 
Let (A,B) denote an interpolating solution which approaches a solution A in $x\to-\infty$, and 
approaches a solution 
B in $x\to+\infty$. 
Then there exist 
\begin{equation}
(\mbox{LD}, \mbox{LD}), \hspace{5mm} 
(AdS_3, \mbox{LD}), \hspace{5mm} 
(AdS_3, AdS_3), 
\end{equation}
but (LD, $AdS_3$) is absent, where LD indicates a linear dilaton solution. 
This would imply the following. 
Consider an $AdS_3$ vacuum with a tachyon. 
By turning on the vev of the tachyon, one could obtain another vacuum, but it cannot be a linear 
dilaton background. 
On the other hand, a linear dilaton background can be possibly 
related to both another linear dilaton and 
$AdS_3$ by a non-zero vev of the tachyon. 
If the existence of an interpolating solution would imply a possible endpoint of tachyon condensation, 
the above results seem to suggest that generically an $AdS_3$ background would be likely realized after 
tachyon condensation, compared to a linear dilaton background.

\vspace{1cm}

\section{Stability of $AdS_3$ vacua  \label{stability}}

\vspace{5mm}

It was observed in the previous section that an $AdS_3$ background often appears as a background 
with a non-zero tachyon vev turned on. 
It is well known that, in $AdS$ spaces, a scalar field may be stable even when its mass$^2$ is 
negative. 
In $AdS_3$ case, a field is stable if 
\begin{equation}
m^2\ge -\frac1{R^2}, 
\end{equation}
where $R$ is the radius of the $AdS_3$. 
This bound on mass is known as the Breitenlohner-Freedman bound. 
Therefore, it would be natural to expect that some $AdS_3$ vacua might be stable at least 
perturbatively, since for a small $AdS_3$ background some tachyonic states in the flat space sense 
might be stabilized. 
If such a vacuum exists, it would be a possible endpoint of a tachyon condensation. 
A similar issue of stability was discussed for $AdS_5$ background in \cite{KT1}\cite{KT2}. 
In the $AdS_3$ case, since string theory on this background is solvable, we can determine whether 
such the stabilization mechanism works well. 

\vspace{5mm}

Let us consider bosonic string theory on $AdS_3\times {\bf R}^n\times M$. 
The mass spectrum of this theory was investigated in \cite{spectrum}. 
A state whose mass$^2$ is below the BF bound corresponds to a state in a principal 
continuous representation 
without spectral flow. 
Let $|j,N\rangle\otimes|p\rangle\otimes|h\rangle$ be a state in the bosonic string theory on 
$AdS_3\times {\bf R}^n\times M$. 
For the principal continuous representation $j=\frac12+is$ with $s$ real, and a non-negative integer 
$N$ represents the contribution to the $L_0$ eigenvalue from non-zero modes of the current algebra. 
$|p\rangle$ is a momentum state in ${\bf R}^n$, and 
$|h\rangle$ is a state of $M$ with the weight $h$. 
The on-shell condition for this state is 
\begin{equation}
\frac1{k-2}\Bigl( \frac14+s^2 \Bigr)+N+\frac{p^2}4+h-1=0, 
   \label{on-shell}
\end{equation}
where $k$ is the level of $SL(2,{\bf R})$ WZW model. 
This implies that $h$ must satisfy 
\begin{equation}
h\le 1-\frac1{4(k-2)}, 
\end{equation}
so that the on-shell condition (\ref{on-shell}) has a solution for some $s,p$ and $N$. 
The level $k$ is determined by imposing that the total central charge is 26, 
\begin{equation}
\frac{3k}{k-2}+n+c_M = 26, 
\end{equation}
where $c_M$ is the central charge of $M$. 
Then the condition for the state to be tachyonic is 
\begin{equation}
h\le \frac{1+n+c_M}{24}. 
   \label{tachyon}
\end{equation}
Since the bulk tachyon is not localized in $M$, by definition, 
the corresponding vertex operator is the unit operator, 
so $h=0$. 
Therefore, no matter how small the radius of $AdS_3$ is, the bulk tachyon cannot be stabilized by the 
background effect. 

Then the best we can now hope is to stabilize all {\it localized} states in $M$. 
Note that the condition (\ref{tachyon}) shows that the number of tachyonic states tends to be small for 
small $n+c_M$. 
In the following, we set $n=0$ so as to reduce the number of localized tachyons. 

First, let us consider the case where $M$ is described by a single minimal model. 
The central charge $c_m$ and the weights $h_{r,s}$ of primary states are as follows, 
\begin{eqnarray}
c_m &=& 1-\frac{6}{m(m+1)}, \\
h_{r,s} &=& \frac{(r(m+1)-sm)^2-1}{4m(m+1)}, 
\end{eqnarray}
where $m\ge3$, $1\le r\le m-1$ and $1\le s \le m$. 
The smallest non-zero weight is $h_{2,2}$ which satisfies 
\begin{equation}
h_{2,2}-\frac{1+c_m}{24} = \frac1{m(m+1)}-\frac1{12} \le 0,
\end{equation}
where the equality holds for $m=3$. 
Therefore, although some states with $h<1$ are stabilized in this background, there is at least one 
tachyonic state unless $m=3$. 
When $M$ is a product of minimal models, the situation is worse; there is always a localized tachyon. 

Next, we consider the case where $M$ is a group manifold $G$.  
The central charge of the corresponding WZW model with level $k$ is 
\begin{equation}
c = \frac{k\mbox{ dim}G}{k+h(G)}. 
\end{equation}
We follow the notations of \cite{Pol}. 
A primary state $|r\rangle$ of the WZW model is labeled by a representation $r$ of the Lie algebra 
of $G$, and its weight $h_r$ is 
\begin{equation}
h_r = \frac{Q_r}{(k+h(G))\psi^2}. 
\end{equation}
Then the condition for the state $|r\rangle$ to be stable is 
\begin{equation}
24Q_r \ge \psi^2k(\mbox{dim}G+1)+\psi^2h(G). 
\end{equation}
Let us focus on $G=SU(N)$. 
By choosing $\psi^2=2$, the above condition becomes 
\begin{equation}
12Q_r \ge kN^2+N.  
   \label{condition}
\end{equation}
Therefore, if this condition holds for all possible representations (except for the trivial 
representation which is not satisfy (\ref{condition})), then it is concluded that there is no localized 
tachyon in the theory. 

Let us check for the fundamental representation. 
For this case $Q_{r}=N-\frac1N$ in our normalization. 
The condition (\ref{condition}) now becomes 
\begin{equation}
k \le \frac{11N^2-12}{N^3}. 
\end{equation}
The solutions of this inequality are 
\begin{eqnarray}
(N,k) &=& (2,1),(2,2),(2,3),(2,4),(3,1),(3,2),(3,3),(4,1),(4,2), \nonumber \\
& & (5,1),(5,2),(6,1),(7,1),(8,1),(9,1),(10,1). 
\end{eqnarray}
For the other combinations of $(N,k)$ there is at least one localized tachyon. 

For $SU(N)$, the quadratic Casimir $Q_r$ has the minimum value when $r$ is the fundamental 
representation. 
This means that 
if $|\mbox{\bf N}\rangle$ is not tachyonic, all the other primary states are not tachyonic. 
Therefore, for the above combinations of $(N,k)$, there is no localized tachyon in the bosonic string 
on $AdS_3\times SU(N)$. 
Since WZW model with level $k$ contains primary states whose corresponding Young diagrams have at most 
$k$ columns, one can check the absence of the localized tachyon just by inspection, using 
\begin{equation}
Q_r = nN-\frac{n^2}N+\sum_{i=1}^k(a_k)^2-\sum_{j=1}^l(b_j)^2, 
\end{equation}
where the Young diagram $Y$ for $r$ has rows of length $a_1\ge a_2\ge\cdots\ge a_k>0$, and columns of 
length $b_1\ge b_2\ge\cdots\ge b_l>0$. 
$n=\sum_{i=1}^ka_k=\sum_{j=1}^lb_j$ is the number of boxes in $Y$. 

\vspace{5mm}

It has been observed that there are choices of $M$ in which all non-trivial states are stable, due 
to the effect of $AdS_3$ background. 
Such a background of the form $AdS_3\times M$ could be a possible endpoint of a condensation of a 
localized tachyon, in a similar sense as the flat space is an endpoint of a localized 
tachyon condensation in Type 0 string on a non-compact orbifold. 
It is very interesting to find other examples of such backgrounds. 
A coset $G/H$ would be a candidate for $M$.

\vspace{1cm}

\section{Discussion  \label{discuss}}

\vspace{5mm}

We discussed classical solutions of the action (\ref{action}) which would be regarded as a 
simplified model for the low energy effective theory of a string theory which has a tachyon. 
We considered all massless fields in the NS-NS sector, and found $AdS_3$ solutions in addition to the 
linear dilaton solutions. 
We also investigated interpolating solutions which connect among $AdS_3$ solutions and linear dilaton 
solutions. 
An interesting observation is that, requiring $K<0$ everywhere, there is an ``asymmetry'' in the 
interpolating solutions; there exists a solution which approaches an $AdS_3$ solution in $x\to-\infty$ 
and linear dilaton solution in $x\to+\infty$, but there is no solution similar to this one with 
$AdS_3$ solution and linear dilaton solution interchanged. 
This might suggest that a linear dilaton solution could not be an endpoint of a tachyon condensation 
starting from an $AdS_3$ solution. 

We also discussed the stability of the $AdS_3$ solutions. 
We found explicit examples of backgrounds, $AdS_3\times SU(N)$, where bosonic string on this background 
does not have tachyons except for the bulk one. 
Primary states of $SU(N)$ WZW models have mass$^2$ above the BF bound. 

The existence of a solution interpolating two $AdS_3$ solutions might suggest a holographic 
description of a tachyon condensation similar to the $AdS$/CFT correspondence \cite{AdS/CFT}. 
Note that the system we discussed only contains the NS-NS fields, so the analysis of classical 
solutions in Type 0 string is parallel to the ones in previous sections. 
The spatial coordinate $x$ is related to $r$ in the $AdS_3$ solution (\ref{AdSmetric}) as 
$r\propto e^{-\frac12K(\pm\infty)x}$. 
This implies that $x\to-\infty$ corresponds to $r\to0$, and $x\to+\infty$ corresponds to $r\to\infty$, 
provided $K<0$ everywhere. 
In other words, the IR limit corresponds to the minimum of $V(T)$, and the UV limit corresponds to 
the maximum of $V(T)$, so the relation between the turning on the tachyon vev and the RG flow 
in terms of $AdS$/CFT is natural. 
If $K(-\infty)$ is chosen to be positive, then $x\to-\infty$ also corresponds to the UV limit. 

A possible existence of a holographic description of tachyon condensation would be a useful tool to 
investigate closed string tachyon condensation. 
In particular, the boundary theory of the string theory on $AdS_3\times SU(N)$, if exists, would be 
useful to investigate the bulk tachyon and its condensation, since the only instability of this 
boundary theory should be related to 
the bulk tachyon, and the analysis of the (im)possibility of its stabilization would be likely 
tractable. 

Since $AdS_3$ background can stabilize all relevant operators in some cases, it would be possible to 
construct an $AdS_3$ vacuum solution of a string theory which is non-supersymmetric but 
perturbatively stable, if the string theory does not have bulk tachyon from the start. 
It would be interesting to discuss its stability in the non-perturbative sense, and also its relation 
to the supersymmetric $AdS_3$ solutions, for example solutions in 
\cite{gaugedSUGRA}\cite{gaugedSUGRA2}.

\vspace{5mm}

\begin{flushleft}
{\Large \bf Acknowledgments}
\end{flushleft}

\vspace{5mm}

I would like to thank M.Fukuma, 
S.Nakamura and T.Takayanagi for valuable discussions. 
I would also like to thank the Yukawa Institute for Theoretical Physics at Kyoto University. 
Discussions during the YITP workshop YITP-W-06-11 on ``String Theory and Quantum Field Theory'' were 
useful to complete this work.
This work was supported in part by JSPS Research Fellowships for Young Scientists.

\end{document}